\documentclass[12pt]{article}
\usepackage{graphics}

\def\bn {\begin{eqnarray}}
\def\en {\end{eqnarray}}
\newcommand{\be}{\begin{equation}}
\newcommand{\ee}{\end{equation}}
\newcommand{\ba}{\begin{eqnarray}}
\newcommand{\ea}{\end{eqnarray}}

\newread\epsffilein    % file to \read
\newif\ifepsffileok    % continue looking for the bounding box?
\newif\ifepsfbbfound   % success?
\newif\ifepsfverbose   % report what you're making?
\newdimen\epsfxsize    % horizontal size after scaling
\newdimen\epsfysize    % vertical size after scaling
\newdimen\epsftsize    % horizontal size before scaling
\newdimen\epsfrsize    % vertical size before scaling
\newdimen\epsftmp      % register for arithmetic manipulation
\newdimen\pspoints     % conversion factor
\def\epsfbox#1{\global\def\epsfllx{72}\global\def\epsflly{72}%
   \global\def\epsfurx{540}\global\def\epsfury{720}%
   \def\lbracket{[}\def\testit{#1}\ifx\testit\lbracket
   \let\next=\epsfgetlitbb\else\let\next=\epsfnormal\fi\next{#1}}%
\def\epsfgetlitbb#1#2 #3 #4 #5]#6{\epsfgrab #2 #3 #4 #5 .\\%
   \epsfsetgraph{#6}}%
\def\epsfnormal#1{\epsfgetbb{#1}\epsfsetgraph{#1}}%
\def\epsfgetbb#1{%
\openin\epsffilein=#1 \ifeof\epsffilein\errmessage{I couldn't open
#1, will ignore it}\else
   {\epsffileoktrue \chardef\other=12

    \def\do##1{\catcode`##1=\other}\dospecials \catcode`\ =10
    \loop
       \read\epsffilein to \epsffileline
       \ifeof\epsffilein\epsffileokfalse\else
          \expandafter\epsfaux\epsffileline:. \\%
       \fi
   \ifepsffileok\repeat
   \ifepsfbbfound\else
    \ifepsfverbose\message{No bounding box comment in #1; using defaults}\fi\fi
   }\closein\epsffilein\fi}%
\def\epsfsetgraph#1{%
   \epsfrsize=\epsfury\pspoints

   \advance\epsfrsize by-\epsflly\pspoints
   \epsftsize=\epsfurx\pspoints
   \advance\epsftsize by-\epsfllx\pspoints
   \epsfxsize\epsfsize\epsftsize\epsfrsize
   \ifnum\epsfxsize=0 \ifnum\epsfysize=0
      \epsfxsize=\epsftsize \epsfysize=\epsfrsize
     \else\epsftmp=\epsftsize \divide\epsftmp\epsfrsize
       \epsfxsize=\epsfysize \multiply\epsfxsize\epsftmp
       \multiply\epsftmp\epsfrsize \advance\epsftsize-\epsftmp
       \epsftmp=\epsfysize
       \loop \advance\epsftsize\epsftsize \divide\epsftmp 2
       \ifnum\epsftmp>0
          \ifnum\epsftsize<\epsfrsize\else
             \advance\epsftsize-\epsfrsize \advance\epsfxsize\epsftmp \fi
       \repeat
     \fi
   \else\epsftmp=\epsfrsize \divide\epsftmp\epsftsize
     \epsfysize=\epsfxsize \multiply\epsfysize\epsftmp
     \multiply\epsftmp\epsftsize \advance\epsfrsize-\epsftmp
     \epsftmp=\epsfxsize
     \loop \advance\epsfrsize\epsfrsize \divide\epsftmp 2
     \ifnum\epsftmp>0
        \ifnum\epsfrsize<\epsftsize\else
           \advance\epsfrsize-\epsftsize \advance\epsfysize\epsftmp \fi
     \repeat
   \fi
   \ifepsfverbose\message{#1: width=\the\epsfxsize, height=\the\epsfysize}\fi
   \epsftmp=10\epsfxsize \divide\epsftmp\pspoints
   \vbox to\epsfysize{\vfil\hbox to\epsfxsize{%
      \includegraphics{#1}%
      \hfil}}%
\epsfxsize=0pt\epsfysize=0pt}%

{\catcode`\%=12 \global\let\epsfpercent=%\global\def\epsfbblit{%BoundingBox}}%
\long\def\epsfaux#1#2:#3\\{\ifx#1\epsfpercent
   \def\testit{#2}\ifx\testit\epsfbblit
      \epsfgrab #3 . . . \\%
      \epsffileokfalse
      \global\epsfbbfoundtrue
   \fi\else\ifx#1\par\else\epsffileokfalse\fi\fi}%
\def\epsfgrab #1 #2 #3 #4 #5\\{%
   \global\def\epsfllx{#1}\ifx\epsfllx\empty
      \epsfgrab #2 #3 #4 #5 .\\\else
   \global\def\epsflly{#2}%
   \global\def\epsfurx{#3}\global\def\epsfury{#4}\fi}%
\def\epsfsize#1#2{\epsfxsize}
\topmargin -0.6cm \textheight 22cm \textwidth  14.5cm
\evensidemargin 5mm \oddsidemargin  10mm
\begin{document}
\begin{center}
{\bf\Large{ A Central Difference Numerical Scheme for Fractional
Optimal Control Problems }}
\end{center}
\begin{center}

{\bf Dumitru Baleanu}\footnote[1]{On leave of absence from
Institute of Space Sciences, P.O.BOX, MG-23, R 76900,
Magurele-Bucharest, Romania, E-mails: dumitru@cankaya.edu.tr,
baleanu@venus.nipne.ro}

Department of Mathematics and Computer Sciences, Faculty of Arts
and Sciences, \c{C}ankaya University, 06530, Ankara, Turkey

{\bf Ozlem Defterli}\footnote[2]{E-mail: defterli@cankaya.edu.tr}
\\ Department of Mathematics and Computer Sciences, Faculty of Arts
and Sciences, \c{C}ankaya University, 06530, Ankara, Turkey

{\bf Om P. Agrawal}\footnote[3]{E-mail:om@engr.siu.edu}
\\ Mechanical Engineering, Southern Illinois University, Carbondale,
Illinois, USA
\end{center}

\begin{abstract}

This paper presents a modified numerical scheme for a class of
Fractional Optimal Control Problems (FOCPs) formulated in Agrawal
(2004) where a Fractional Derivative (FD) is defined in the
Riemann-Liouville sense. In this scheme, the entire time domain is
divided into several sub-domains, and a fractional derivative
(FDs) at a time node point is approximated using a modified
Gr\"{u}nwald-Letnikov approach. For the first order derivative,
the proposed modified Gr\"{u}nwald-Letnikov definition leads to a
central difference scheme. When the approximations are substituted
into the Fractional Optimal Control (FCO) equations, it leads to a
set of algebraic equations which are solved using a direct
numerical technique. Two examples, one time-invariant and the
other time-variant, are considered to study the performance of the
numerical scheme. Results show that 1) as the order of the
derivative approaches an integer value, these formulations lead to
solutions for integer order system, and 2) as the sizes of the
sub-domains are reduced, the solutions converge.  It is hoped that
the present scheme would lead to stable numerical methods for
fractional differential equations and optimal control problems.
\end{abstract}


\section{Introduction}

Optimal Control Problems (OCPs) appear in engineering, science,
economics, and many other fields.  An extensive body of work
exists in the area of optimal control of integer order dynamic
systems ( Hestenes (1966), Bryson and Ho(1975), Gregory and
Lin(1992)). It was shown recently that fractional derivatives
provide more accurate behavior of a dynamic system (see Podlubny
(1999) and the references there in). Therefore, formulations and
numerical schemes for optimal control problems which account for
fractional dynamics of these systems would be necessary. In this
work, we develop a modified numerical scheme for a class of
Fractional Optimal Control Problems whose dynamics is described by
Fractional Differential Equations.

Agrawal (Agrawal (2004)) defines a {\it Fractional Dynamic System}
(FDS) as a system whose dynamics is described by {\it Fractional
Differential Equations} (FDEs), and a {\it Fractional Optimal
Control Problem} (FOCP) as an optimal control problem for an FDS.
A general formulation for FOCPs was proposed in Agrawal (2004). As
it can be seen from literature, there is no much work in the field
of optimal control of FDSs. The formulations of FOCPs comes from
Fractional Variational Calculus (FVC) which is an emerging branch
of fractional calculus.

Riewe (Riewe (1996), Riewe (1997)) was first to formulate a
fractional variational mechanics problem. Riewe's major focus was
to develop Lagrangian and Hamiltonian mechanics for dissipative
systems. Agrawal (Agrawal (2001)) presented an ad hoc approach to
obtain the differential equations of fractionally damped systems.
In Agrawal (2002), Agrawal presented fractional Euler-Lagrange
equations for Fractional Variational Problems (FVPs).  Klimek
(Klimek (2001)) presented a fractional sequential mechanics model
with symmetric fractional derivatives. In Klimek (2002), Klimek
presented stationary conservation laws for fractional differential
equations with variable coefficients. Dreisigmeyer and Young
(2003) presented nonconservative Lagrangian mechanics using a
generalized function approach. In Dreisigmeyer and Young(2004) the
authors show that obtaining differential equations for a
nonconservative system using FDs may not be possible.

The fractional Euler-Lagrange equation has recently been used by
Baleanu and coworker to model fractional Lagrangian with linear
velocities (Baleanu and Avkar(2004)), fractional metafluid
dynamics (Baleanu (2004)), fractional Lagrangian and Hamiltonian
formulations of discrete and continuous systems (Muslih and
Baleanu (2005a), Muslih and Baleanu (2005b), Muslih et al.(2006),
Baleanu and Muslih(2005)) and Hamiltonian analysis of irregular
systems (Baleanu (2006)). Tarasov and Zaslavsky have used
variational Euler-Lagrange equation to derive fractional
generalization of the Ginzburg-Landau equation for fractal media
(Tarasov and Zaslavsky(2005)) and dynamic systems subjected to
nonholonomic constraints (Tarasov and Zaslavsky(2006)). In
(Agrawal (2004), Agrawal (2005)), the fractional variational
calculus is applied to deterministic and stochastic analysis of
fractional optimal control problems.  Rabei, Ajlouni and Ghassib
(2006) develop suitable Lagrangian and Hamiltonian for a
fractional dynamic system, which they transform to fractional
Schrodinger's equation and solve it. Stanislavsky (2006) presents
a Hamiltonian formulation of a dynamic system.  Atanackovic and
Stankovic (2007) present existence and uniqueness criteria for
problems resulting from fractional variational calculus.

In this paper, we present a direct numerical scheme for a class of
Fractional Optimal Control Problems (FOCPs) formulated in Agrawal
(2004).  The scheme uses a modified Gr\"{u}nwald-Letnikov
definition to approximate a fractional derivative. For a first
order derivative, this approximation leads to a central difference
formula.  For simplicity in the discussion to follow, this
formulation is briefly presented here.  Two examples are solved to
demonstrate the performance of the algorithm.

\section{Fractional Optimal Control Formulation}

In this section, we briefly present a Hamiltonian formulation for
an FOCP. Consider the following FOCP: Find the optimal control
$u(t)$ for a FDS that minimizes the performance index
\begin{equation}\label{eq6}
J(u) = \int_0^1 f(x, u, t) dt
\end{equation}
and satisfies the system dynamic constraints
\begin{equation}\label{eq7}
_0D_t^\alpha x = g(x, u, t),
\end{equation}
and the initial condition
\begin{equation}\label{eq8}
x(0) = x_0 ,
\end{equation}
where $x(t)$ is the state variable, $t$ represents the time, $f$
and $g$ are two arbitrary functions, and $_0D_t^\alpha x$
represents the left Riemann-Liouville derivative of order $\alpha$
of $x$ with respect to $t$.  For the definitions of fractional
derivatives and some of their applications, see (Podlubny (1999),
Magin (2006), and Kilbas, Srivastava and Trujillo (2006)).  Note
that the upper limit of the integration is taken as 1.  We
consider $0<\alpha<1$. Further, we consider that $x(t)$, $u(t)$,
$f(x,u,t)$  and $g(x,u,t)$ are all scalar functions.  These
conditions are made for simplicity.  The same procedure could be
followed if the upper limit of integration and $\alpha$ are
greater than 1, and $x(t)$, $u(t)$, $f(x,u,t)$ and $g(x,u,t)$ are
vector functions.

It should be pointed out that in traditional integer-order optimal
Control, Eq. (1) may also include terminal terms.  Such terms lead
to nonzero terminal condition at $t=1$.  For the FOCP considered
here, our formulation would require fractional terminal terms, the
meaning of which may not be clear.  For this reason, the terminal
terms are not included in Eq. (1).

To find the optimal control we define a modified performance index
as
\begin{equation}\label{eq9}
\bar{J}(u) = \int_0^1 [H(x, u, t) - \lambda \,_0D_t^\alpha x ] dt,
\end{equation}
where $H(x,u,\lambda, t)$ is the Hamiltonian of the system defined
as
\begin{equation}\label{eq10}
H(x, u, \lambda, t) = f(x,u,t) + \lambda g(x,u,t),
\end{equation}
and $\lambda$ is the Lagrange multiplier.  Taking variations of
Eq. (\ref{eq9}) and using (\ref{eq10}), the necessary equations
for the optimal control are given as
\begin{equation}\label{eq11}
_tD_1^\alpha \lambda = \frac{\partial H}{\partial x},
\end{equation}
\begin{equation}\label{eq12}
\frac{\partial H}{\partial u} = 0,
\end{equation}
and
\begin{equation}\label{eq13}
_0D_t^\alpha x = \frac{\partial H}{\partial \lambda}
\end{equation}
Following the approach presented in Agrawal (2004), we also
require that
\begin{equation}\label{eq14}
\lambda(1) = 0.
\end{equation}
Equations (\ref{eq11})-(\ref{eq14}) represent the necessary
conditions in terms of a Hamiltonian for the optimal control of
the FOCP defined above.  It could be verified that the total time
derivative of the Hamiltonian as defined above is not zero along
the optimum trajectory even when $f$ and $g$ do not explicitly
depend on $t$.  This is a departure from the integer order optimal
control theory.

In the discussion to follow, we shall strictly focus on the
following quadratic performance index
\begin{equation}\label{eq15}
J(u) = \frac{1}{2} \int_0^1 [ q(t) x^2(t) + r(t) u^2 ] dt ,
\end{equation}
where $q(t) \ge 0$ and $r(t) > 0$, and the system whose dynamics
is described by the following linear FDE,
\begin{equation}\label{eq16}
_0D_t^\alpha x = a(t) x + b(t) u .
\end{equation}
Using Eqs. (\ref{eq11}) to (\ref{eq13}), it can be demonstrated
that the necessary Euler-Lagrange equations for this system are
(see also Agrawal (2004)),
\begin{equation}\label{eq17}
_0D_t^\alpha x = a(t) x - r^{-1}(t) b^2(t) \lambda,
\end{equation}
\begin{equation}\label{eq18}
_tD_1^\alpha \lambda = q(t) x + a(t) \lambda,
\end{equation}
and
\begin{equation}\label{eq19}
u = - r^{-1}(t) b(t) \lambda.
\end{equation}
Equations (\ref{eq17}) to (\ref{eq19}) will be used to develop a
direct numerical scheme for a FOCP.

\section{A Modified Numerical Scheme for FOCPs}

In this section, we define modified Gr\"{u}nwald-Letnikov
approximations of fractional derivatives as
\begin{equation}\label{ecuu1}
{}_0\textbf{D}_t^{\alpha}x(t_{i-1/2})\cong \frac{1}{h^{\alpha}}
\sum_{j=0}^i\omega_j^{(\alpha)} x_{i-j}, \hspace{0.2in} i=1,
\cdots, n.
\end{equation}
\begin{equation}\label{ecuu2}
{}_t\textbf{D}_1^{\alpha}u(t_{i+1/2})\cong \frac{1}{h^{\alpha}}
\sum_{j=0}^{n-i}\omega_j^{(\alpha)}u_{i+j}, \hspace{0.2in} i=n-1,
n-2, \cdots, 0,
\end{equation}
where $\omega_j^{(\alpha)}$, $ j=0, 1,\cdots, n$ are the
coefficients. A recursive approach of computing
$\omega_j^{(\alpha)}$ is given as (Podlubny (1999))
$$ \omega_0^{(\alpha)} = 1, \hspace{0.2in} \omega_j^{(\alpha)} =
\left( 1–-\frac{\alpha+1}{j} \right) \omega_{j-1}^{(\alpha)},
\hspace{0.2in} j=1,\cdots, n. $$ It can be shown that for
$\alpha=1$, Eqs. (\ref{ecuu1}) and (\ref{ecuu2}) lead to
$$ \frac{d x (t_{i-1/2})}{dt} = \frac{x_i-–x_{i-1}}{h} $$
and
$$ - \frac{d x (t_{i+1/2})}{dt} = \frac{x_i–-x_{i+1}}{h} $$
which are essentially the central difference equations for the
left and the right derivatives.

To develop a numerical scheme, we divide the time domain [0, 1]
into $n$ equal parts, and approximate the fractional derivatives
$_0D_t^\alpha x$ and $_tD_1^\alpha \lambda$ at the center of each
part using Eqs. (\ref{ecuu1}) and (\ref{ecuu2}).  We further take
$x(t_{i-1/2})$ as an average of the two end values of the segment.
Thus, $x(t_{i-1/2})= (x_{i-1}+x_i)/2$.  We make similar
approximations for $\lambda(t_{i-1/2})$, $x(t_{i+1/2})$, and
$\lambda(t_{i+1/2})$. Substituting these approximations into
(\ref{eq17}) and (\ref{eq18}), we obtain
$$ \frac{1}{h^{\alpha}}\sum_{j=0}^i\omega_j^{(\alpha)} x_{i-j} =
\frac{1}{2} a( i_1 h) (x_{i-1}+x_i)- \frac{1}{2} r^{-1} (i_1 h)
b^2(i_1 h) (\lambda_{i-1}+ \lambda_i) $$
\begin{eqnarray}\label{eq17a}
\hspace{3.0in} i=1, \cdots, n
\end{eqnarray}
$$\frac{1}{h^{\alpha}}\sum_{j=0}^{n-i}\omega_j^{(\alpha)}
\lambda_{i+j}= \frac{1}{2} q( i_2 h) (x_{i+1}+x_i) + \frac{1}{2}
a( i_2 h) (\lambda_{i-1}+ \lambda_i) $$
\begin{equation}\label{eq18a}
\hspace{3.0in} i=n-1, \cdots, 0
\end{equation}
where $i_1 = i-\frac{1}{2}$ and $i_2 = i+\frac{1}{2}$.  Equations
(\ref{eq17a}) and (\ref{eq18a}) represent a set of $2n$ linear
equations in terms of $2n$ unknowns, which can be solved using a
standard linear solver.  One can also develop an iterative scheme
in which one can march forward to compute $x_i$'s and backward to
compute $\lambda_i$'s to save storage space and perhaps
computational time.

%As a result we obtain

%\begin{equation}
%\sum_{j=0}^i\omega_j^{(\alpha)}
%x_{i-j}+\frac{h^{\alpha}}{2}(x_{i-1}+x_i)-\frac{h^{\alpha}}{2}(u_{i-1}+u_i)=0,i=1,\cdots,
%n.
%\end{equation}

%\begin{equation}
%\sum_{j=0}^{n-i}\omega_j^{(\alpha)}u_{i+j}+
%\frac{h^{\alpha}}{2}(x_i+x_{i+1}) + \frac{h^\alpha}{2}(u_i
%+u_{1+i})=0, i=n-1, n-2,\cdots,0.
%\end{equation}

\section{Numerical Examples}

   To demonstrate the applicability of the formulation and to
validate the numerical scheme, in this section we present
numerical results for two problems, time invariant and time
varying.  For both problems, two types of studies were conducted.
The first study involved examination of the response as the number
of divisions was increased. For this purpose, $N$ was taken as 8,
16, 32, 64, 128 and 256.  The second study involved examination of
the response as the order of derivatives approach 1.  Results of
these studies are given below.

\subsection{Time Invariant FOCP}

As a first example, we consider the following Time Invariant
Problem (TIP): Find the control $u(t)$ which minimizes the
quadratic performance index
\begin{equation}\label{eq23}
J(u) = \frac{1}{2} \int_0^1 [ x^2(t) + u^2(t) ] dt
\end{equation}
subjected to the system dynamics
\begin{equation}\label{eq24}
_0D_t^{\alpha} x = - x + u,
\end{equation}
and the initial condition
\begin{equation}\label{eq25}
x(0) = 1.
\end{equation}
For this example, we have
\begin{equation}\label{eq26}
q(t) = r(t) = - a(t) = b(t) = x_0 = 1.
\end{equation}
This example is considered here because, for $\alpha=1$, it is one
of the most common examples of time invariant systems considered
by many.  The closed form solution for this system for $\alpha=1$
is given as (see, Agrawal (1989))
\begin{equation}\label{eq27a}
x(t)=\cosh (\sqrt{2}t)+\beta \sinh (\sqrt{2}t)
\end{equation}
and
\begin{equation}\label{eq27b}
u(t)=(1+\sqrt{2}\beta) \cosh (\sqrt{2}t)+(\sqrt{2}+\beta) \sinh
(\sqrt{2}t)
\end{equation}
where
\begin{equation}\label{eq27c}
\beta = - \frac{\cosh (\sqrt{2})+\sqrt{2}\sinh
(\sqrt{2})}{\sqrt{2} \cosh (\sqrt{2})+\sinh (\sqrt{2})} \approx -
0.9799.
\end{equation}
From Eqs. (\ref{eq27b}) and (\ref{eq27c}), we get $u(0) =
-0.3858$.

Figures 1 and 2 show the state $x(t)$ and the control $u(t)$ as
functions of $t$ for $\alpha=0.75$ and different values of $N$.
%\newpage
\input epsf
\begin{center}
\epsfxsize=9cm \epsfbox{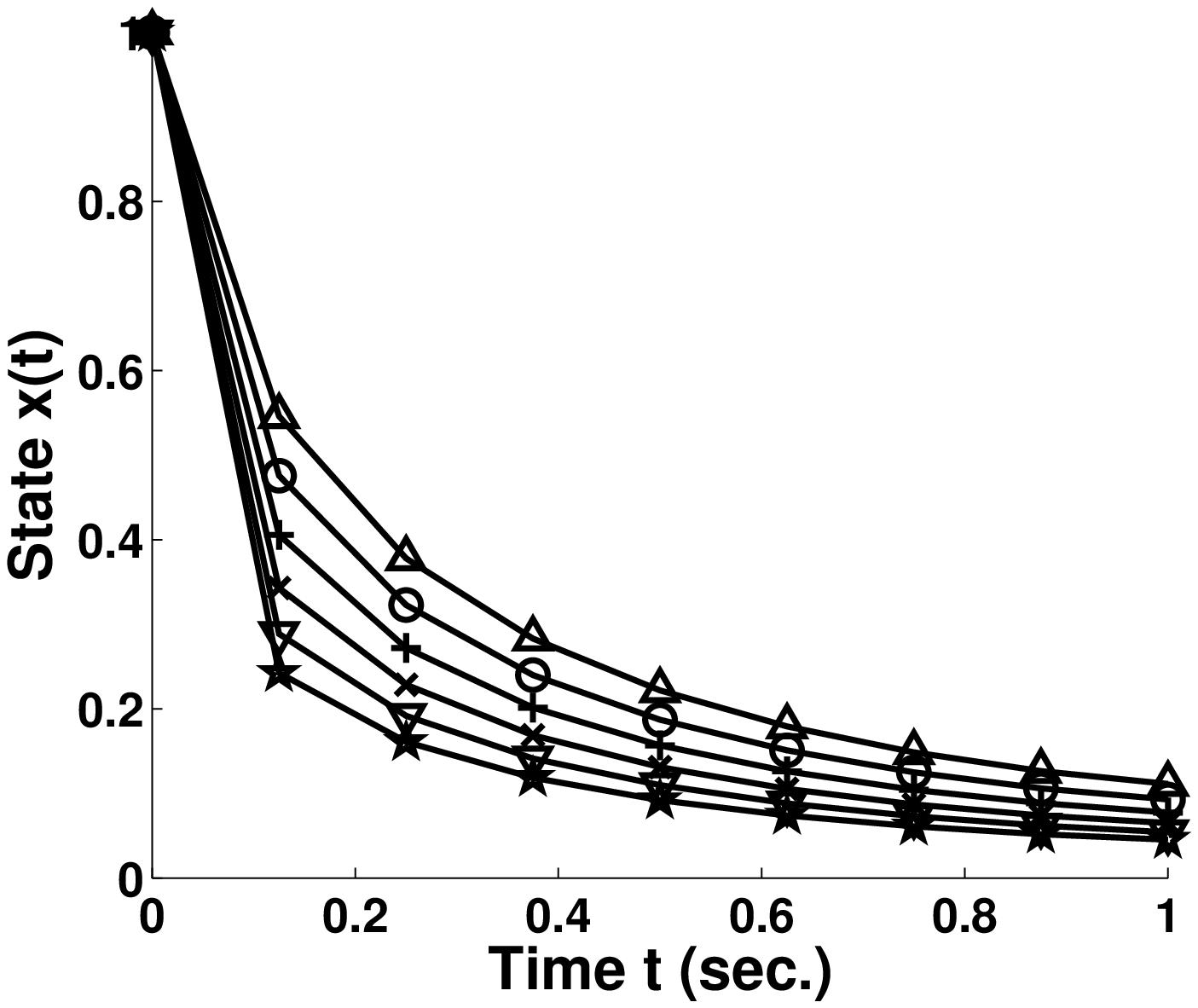}

Figure 1: Convergence of $x(t)$ for the TIP for $\alpha=0.75$
($\Delta:N=8$, $O:N=16$, $+:N=32$, $X:N=64$, $\nabla:N=128$,
$\star:N=256$)
\end{center}
%\begin{figure}
 %\centerline{\includegraphics{Fig1Prob1.eps}}
 %\caption{Convergence of $x(t)$ for the TIP for $\alpha=0.75$
%($\Delta:N=8$, $O:N=16$, $+:N=32$, $X:N=64$, $\nabla:N=128$)}
 %\label{fig1}
%\end{figure}

\begin{center}
\epsfxsize=9cm \epsfbox{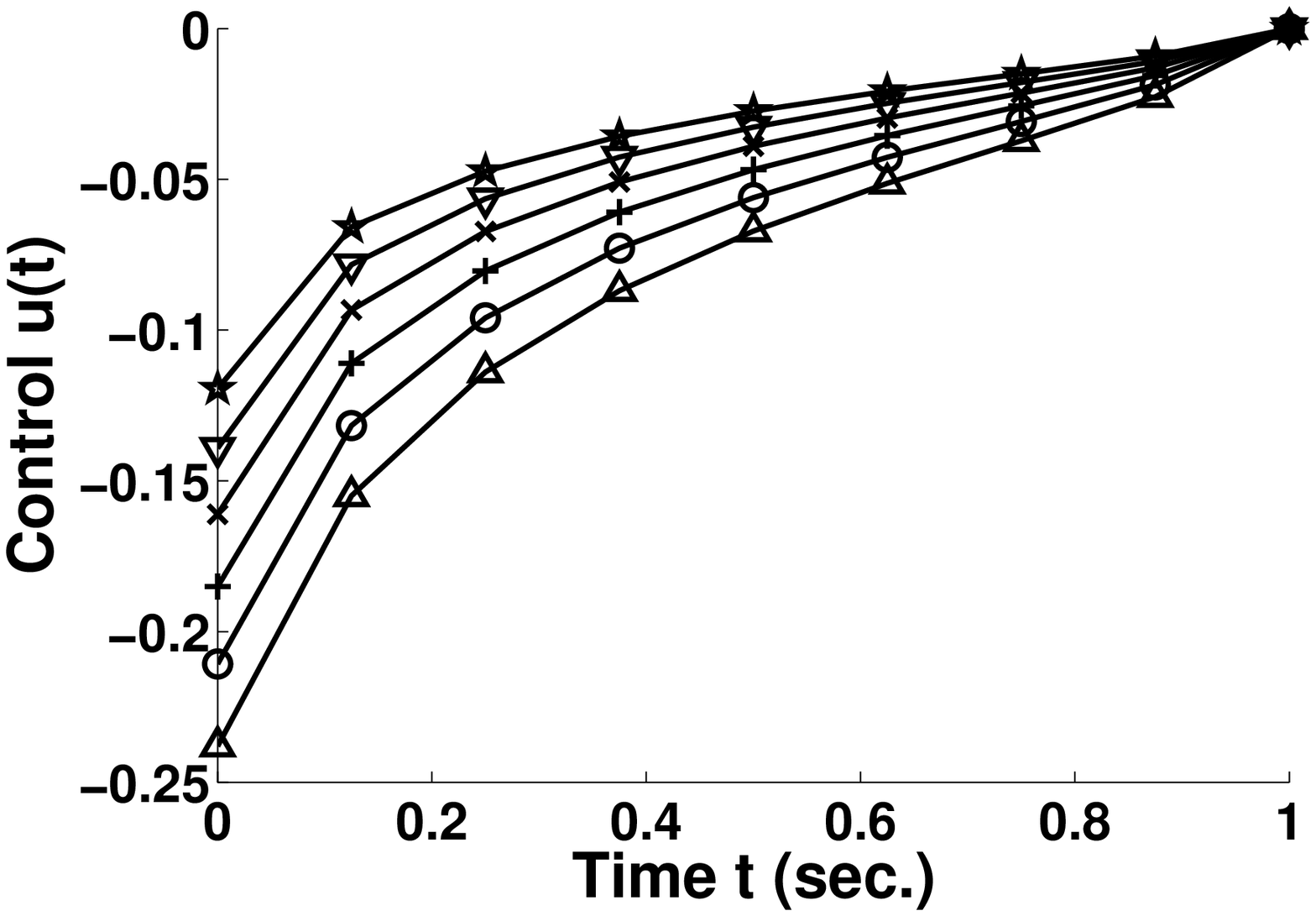}

Figure 2: Convergence of $u(t)$ for the TIP for $\alpha=0.75$
($\Delta:N=8$, $O:N=16$, $+:N=32$, $X:N=64$, $\nabla:N=128$,
$\star:N=256$)
\end{center}

%\begin{figure}
%\begin{figure}[htb]
 %\centerline{\includegraphics{Fig2Prob1.eps}}
 %\caption{Convergence of $u(t)$ for the TIP for $\alpha=0.75$
%($\Delta:N=8$, $O:N=16$, $+:N=32$, $X:N=64$, $\nabla:N=128$)}
 %\label{fig2}
%\end{figure}

Figures 3 and 4 show the state $x(1)$ and the control $u(0)$ as a
function of $N$ for different $\alpha$.  From these figures, it
can be seen that the solutions converge as $N$ is increased,
however, the convergence is slow.  Further, the convergence
becomes poor as $\alpha$ is decreased.  Further error analysis may
be necessary to identify the reasons for this behavior.

\begin{center}
\epsfxsize=9cm \epsfbox{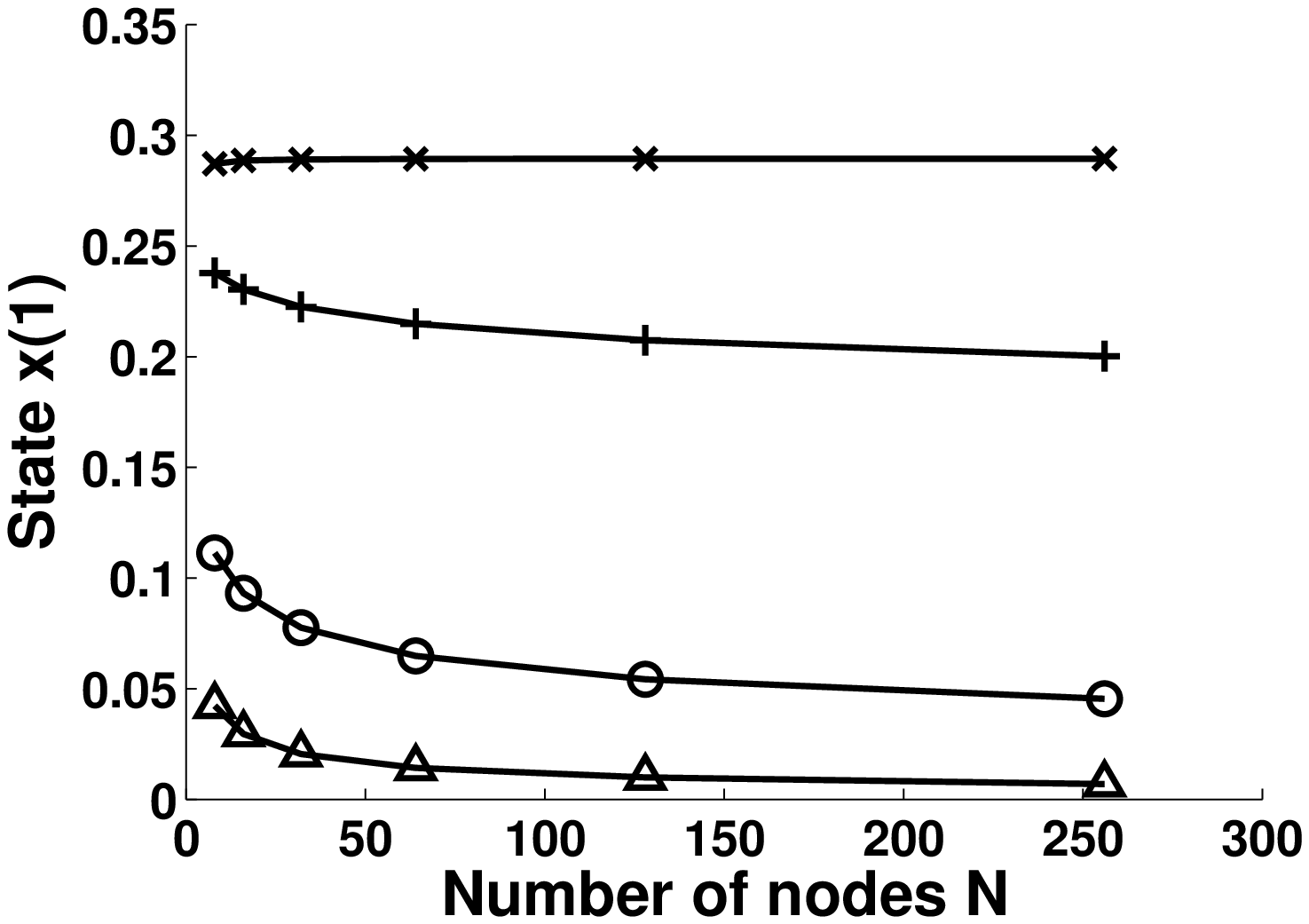}

Figure 3: Convergence of $x(1)$ for the TIP for different $\alpha$
($\Delta:\alpha=0.5$, $O:\alpha=0.75$, $+:\alpha=0.95$,
$X:\alpha=1.0$)
\end{center}

%\begin{figure}
%\begin{figure}[htb]
 %\centerline{\includegraphics{convX1Prob1.eps}}
 %\caption{Convergence of $x(1)$ for the TIP for different $\alpha$
%($\Delta:\alpha=0.5$, $O:\alpha=0.75$, $+:\alpha=0.95$,
%$X:\alpha=1.0$)}
 %\label{fig3}
%\end{figure}

\begin{center}
\epsfxsize=9cm \epsfbox{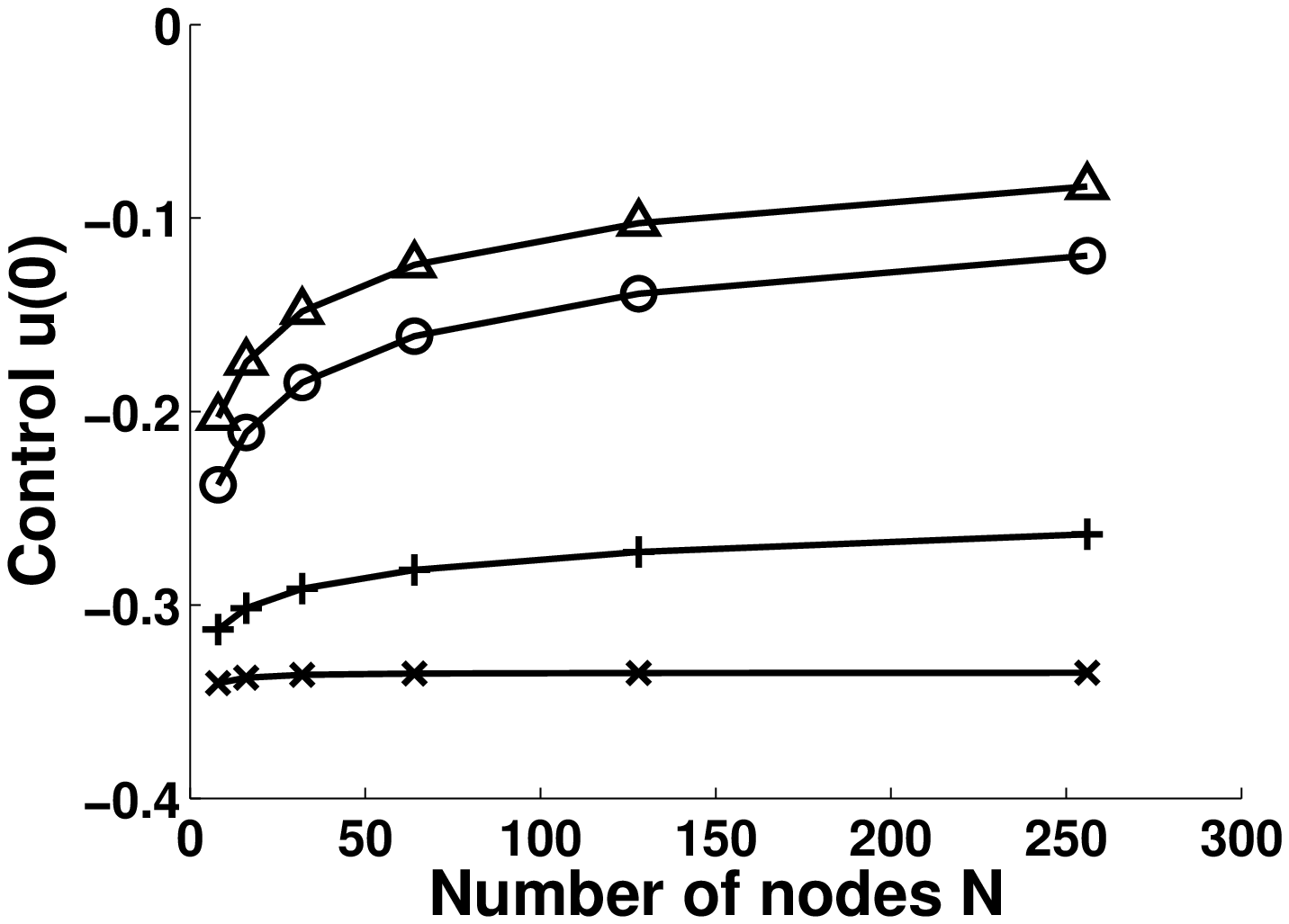}

Figure 4: Convergence of $u(0)$ for the TIP for different $\alpha$
($\Delta:\alpha=0.5$, $O:\alpha=0.75$, $+:\alpha=0.95$,
$X:\alpha=1.0$)
\end{center}

%\begin{figure}
%\begin{figure}[htb]
 %\centerline{\includegraphics{convU1Prob1.eps}}
 %\caption{Convergence of $u(0)$ for the TIP for different $\alpha$
%($\Delta:\alpha=0.5$, $O:\alpha=0.75$, $+:\alpha=0.95$,
%$X:\alpha=1.0$)}
 %\label{fig4}
%\end{figure}
Figures 5 and 6 show the state $x(t)$ and the control $u(t)$ as
functions of $t$ for different values of $\alpha$.  These figures
also show analytical results for the state $x(t)$ and the control
$u(t)$ for $\alpha=1$.  It can be observed that for $\alpha=1$ the
numerical solution agrees with the analytical solution.  Thus, as
$\alpha$ approaches to 1, the solution for the integer order
system is recovered.

\begin{center}
\epsfxsize=9cm \epsfbox{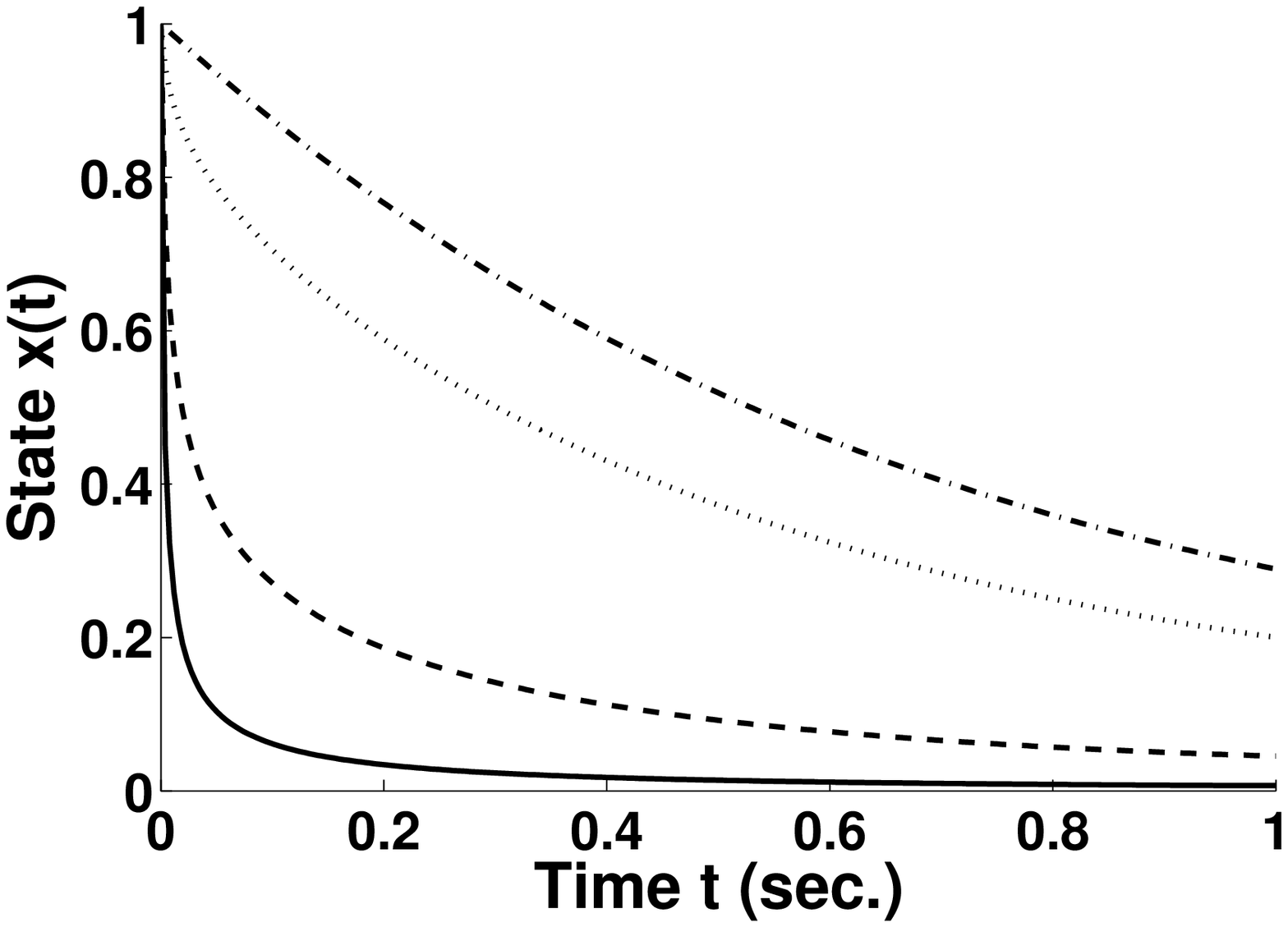}

Figure 5: State $x(t)$ as a function of $t$ for the TIP for
different $\alpha$\\ ( $ -$ :$\alpha=0.5$, -$ $-$ $-
:$\alpha=0.75$, $\cdot$$\cdot$$\cdot$ :$\alpha=0.95$,
-$\cdot$-$\cdot$- :$\alpha=1$)
\end{center}
%\begin{figure}
%\begin{figure}[htb]
 %\centerline{\includegraphics{state256XProb1.eps}}
 %\caption{State $x(t)$ as a function of $t$ for the TIP for
%different $\alpha$ ($\Delta:\alpha=0.5$, $O:\alpha=0.75$,
%$+:\alpha=0.95$, $X:\alpha=1$, $\nabla:\mbox{Analytical for }$
% $\alpha=1$)}
% \label{fig5}
%\end{figure}

\begin{center}
\epsfxsize=9cm \epsfbox{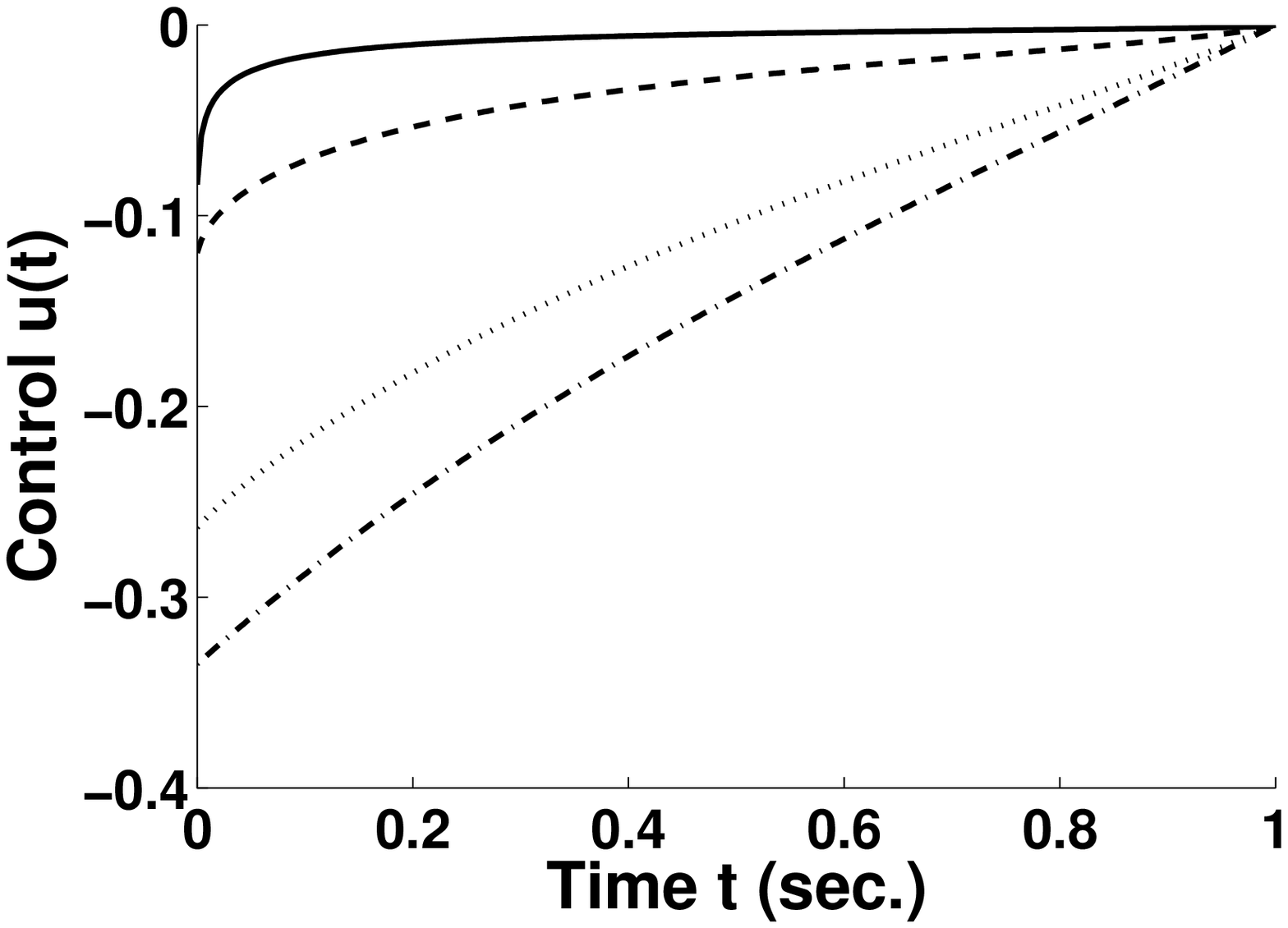}

Figure 6: Control $u(t)$ as a function of $t$ for the TIP for
different $\alpha$\\ ( $ -$ :$\alpha=0.5$, -$ $-$ $-
:$\alpha=0.75$, $\cdot$$\cdot$$\cdot$ :$\alpha=0.95$,
-$\cdot$-$\cdot$- :$\alpha=1$)

\end{center}
%\begin{figure}
%\begin{figure}[htb]
 %\centerline{\includegraphics{state256UProb1.eps}}
 %\caption{Control $u(t)$ as a function of $t$ for the TIP for
%different $\alpha$ ($\Delta:\alpha=0.5$, $O:\alpha=0.75$,
%$+:\alpha=0.95$, $X:\alpha=1$, $\nabla:\mbox{Analytical for }$
% $\alpha=1$)}
 %\label{fig6}
%\end{figure}

\subsection{Time Varying FOCP}

As a second example, we consider the following Time Varying
Problem (TVP): Find the control $u(t)$ which minimizes the
quadratic performance index given in Eq. (\ref{eq23}), and which
satisfies the system dynamics
\begin{equation}\label{eq28}
_0D_t^\alpha x = t x + u.
\end{equation}
The initial condition is $x(0) = 1$.  For this example, we have
\begin{equation}\label{eq30}
q(t) = r(t) = b(t) = x_0 = 1, \hspace{0.2in} a(t) = t.
\end{equation}
It is one of the simplest examples of time varying systems, and
for $\alpha=1$, it has been considered at several other places
(see, Agrawal (1989), and the references there in).

Figures 7 and 8 show the state $x(t)$ and the control $u(t)$ as
functions of $t$ for different values of $N$.  Figures 9 and 10
show the state $x(1)$ and the control $u(0)$ as a function of $N$
for different $\alpha$.  As for the TIP, the solutions for the TVP
also converge as $N$ is increased, however, as before, the
convergence is slow.  This slow convergence for both examples
clearly suggests that further improvement of the scheme is
necessary.  Figures 11 and 12 show the state $x(t)$ and the
control $u(t)$ as functions of $t$ for different values of
$\alpha$.

\begin{center}
\epsfxsize=9cm \epsfbox{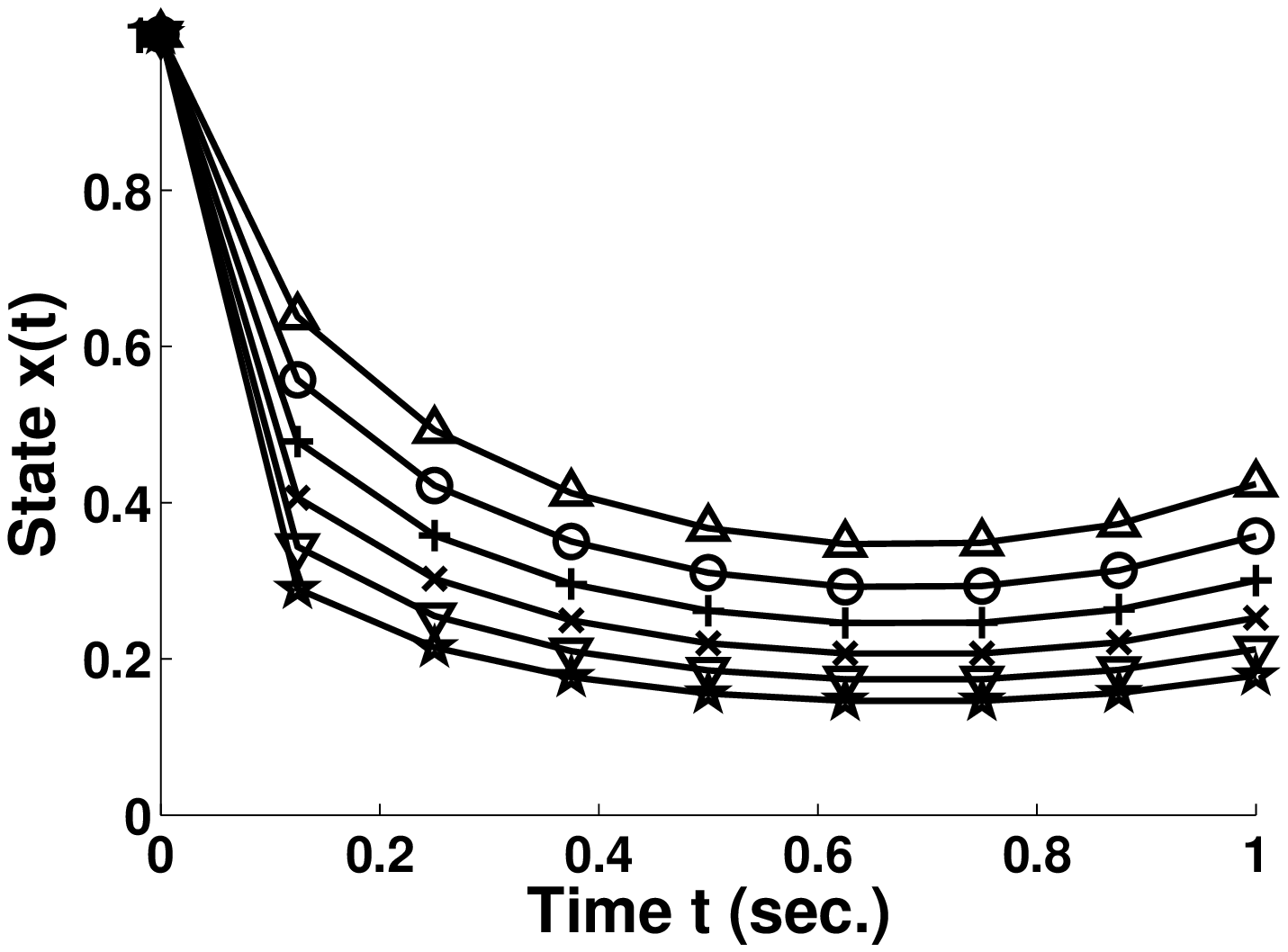}

Figure 7: Convergence of $x(t)$ for the TVP for $\alpha=0.75$
($\Delta:N=8$, $O:N=16$, $+:N=32$, $X:N=64$, $\nabla:N=128$,
$\star:N=256$)
\end{center}

%\begin{figure}
%\begin{figure}[htb]
 %\centerline{\includegraphics{Fig1Prob2.eps}}
  % \caption{Convergence of $x(t)$ for the TVP for $\alpha=0.75$
%($\Delta:N=8$, $O:N=16$, $+:N=32$, $X:N=64$, $\nabla:N=128$)}
 %\label{fig7}
%\end{figure}

\begin{center}
\epsfxsize=9cm \epsfbox{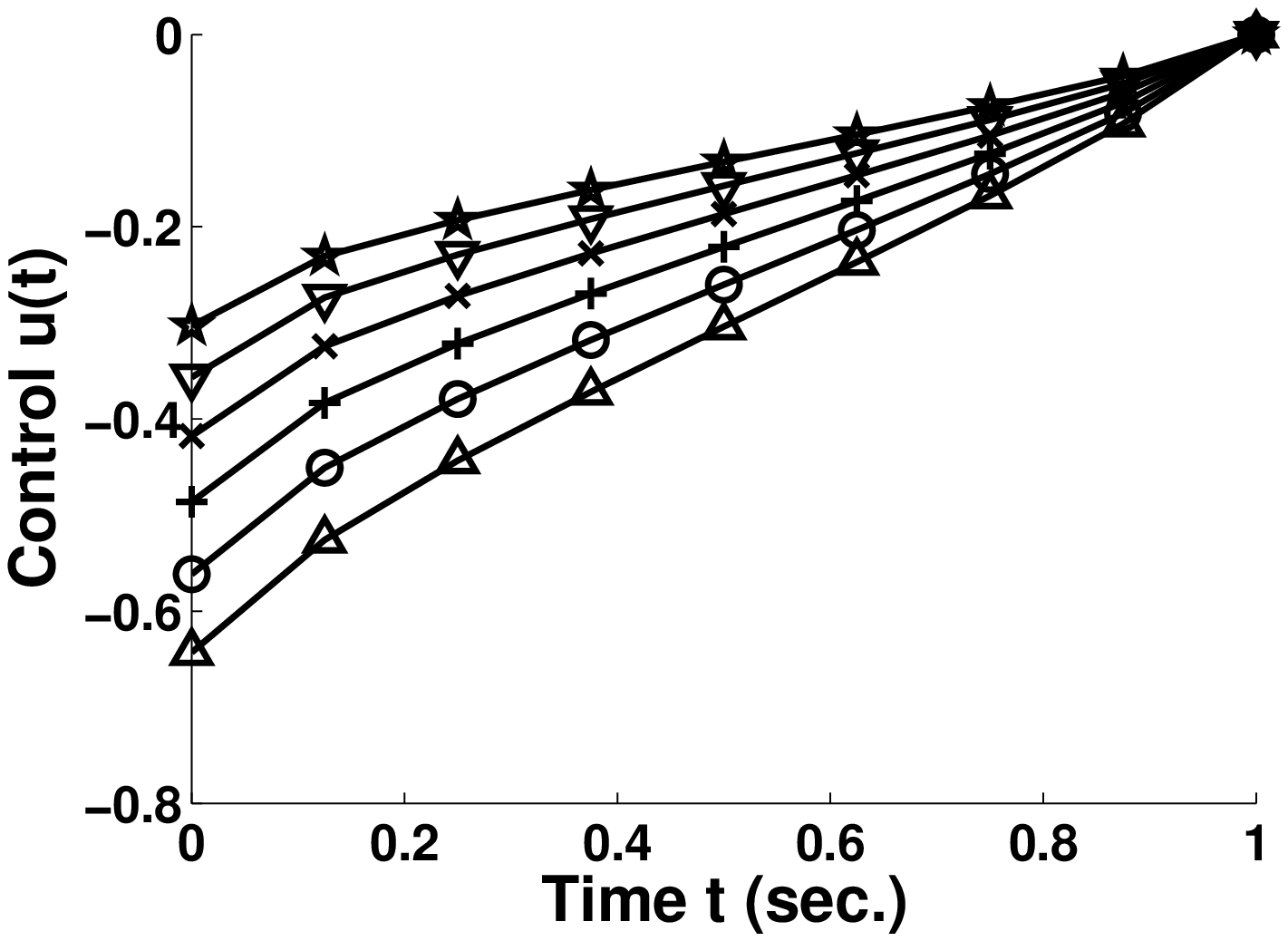}

Figure 8: Convergence of $u(t)$ for the TVP for $\alpha=0.75$
($\Delta:N=8$, $O:N=16$, $+:N=32$, $X:N=64$, $\nabla:N=128$,
$\star:N=256$)
\end{center}

%\begin{figure}
%\begin{figure}[htb]
 %\centerline{\includegraphics{Fig2Prob2.eps}}
 %\caption{Convergence of $u(t)$ for the TVP for $\alpha=0.75$
%($\Delta:N=8$, $O:N=16$, $+:N=32$, $X:N=64$, $\nabla:N=128$)}
 %\label{fig8}
%\end{figure}

%\newpage

\begin{center}
\epsfxsize=9cm \epsfbox{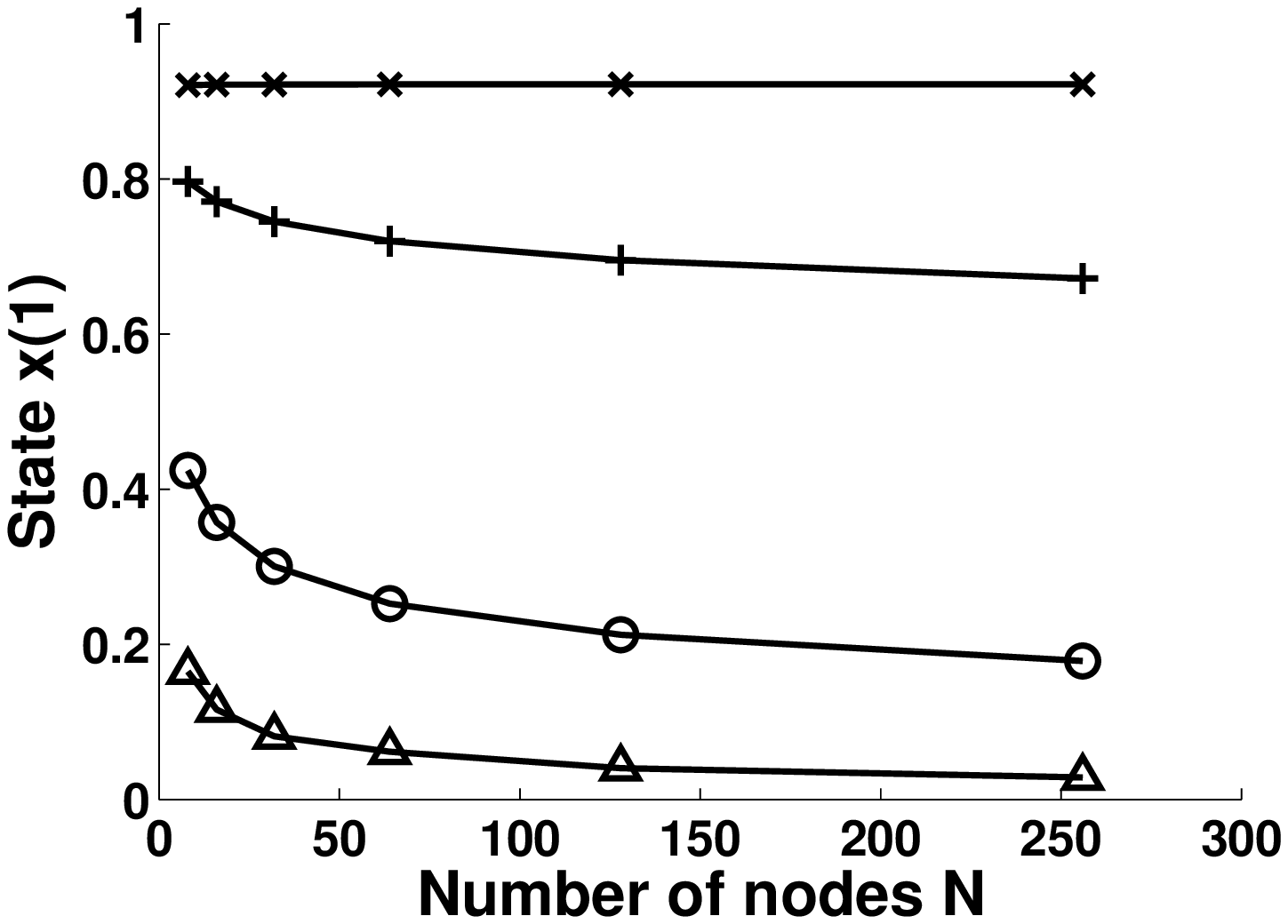}

Figure 9: Convergence of $x(1)$ for the TVP for different $\alpha$
($\Delta:\alpha=0.5$, $O:\alpha=0.75$, $+:\alpha=0.95$,
$X:\alpha=1.0$)
\end{center}

\begin{center}
\epsfxsize=9cm \epsfbox{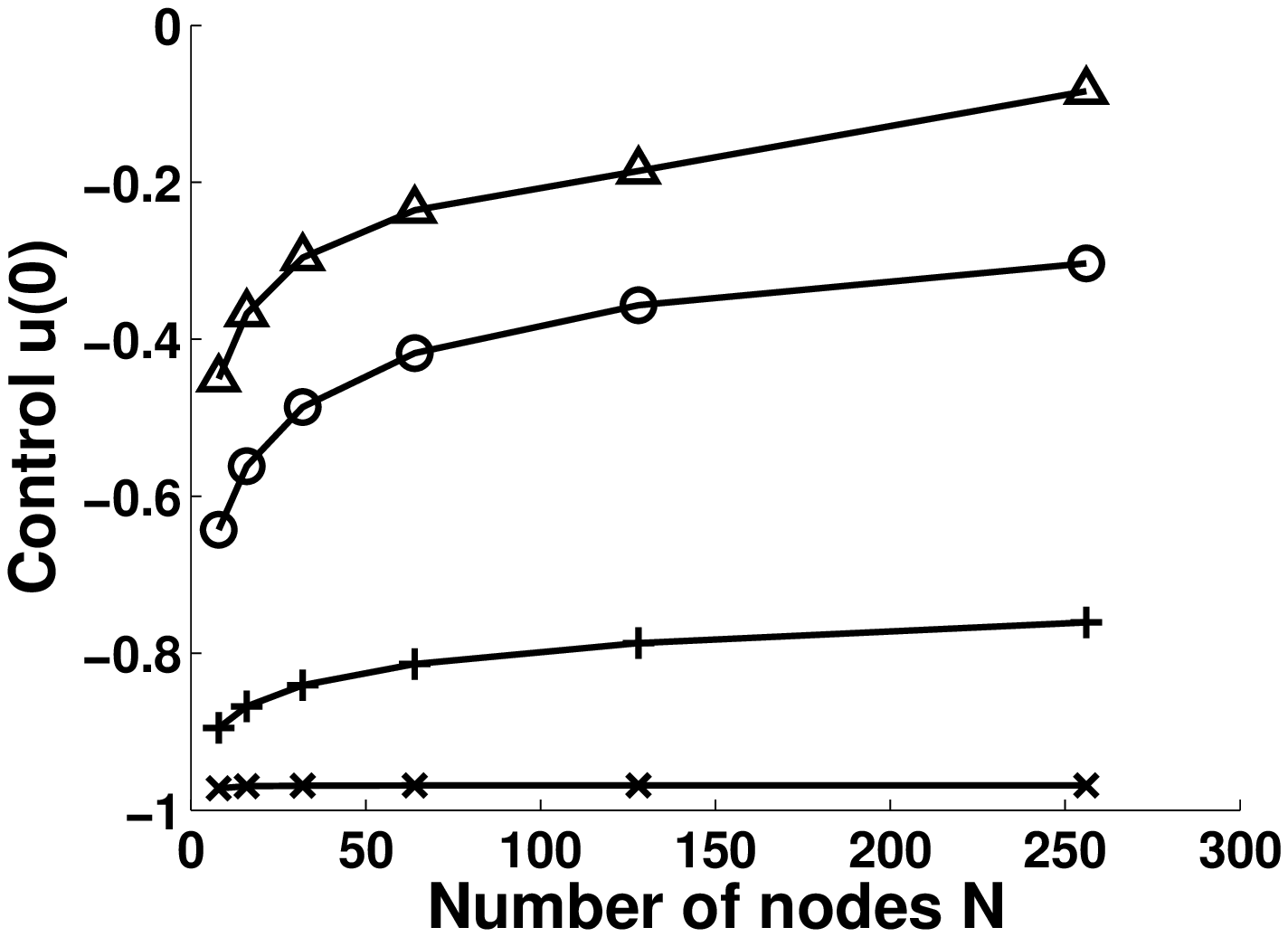}

Figure 10: Convergence of $u(0)$ for the TVP for different
$\alpha$ ($\Delta:\alpha=0.5$, $O:\alpha=0.75$, $+:\alpha=0.95$,
$X:\alpha=1.0$)
\end{center}
%\begin{figure}
%\begin{figure}[htb]
 %\centerline{\includegraphics{convU1Prob2.eps}}
 %\caption{Convergence of $u(0)$ for the TVP for $\alpha=0.75$
%($\Delta:\alpha=0.5$, $O:\alpha=0.75$, $+:\alpha=0.95$,
%$X:\alpha=1.0$)}
% \label{fig10}
%\end{figure}

\begin{center}
\epsfxsize=9cm \epsfbox{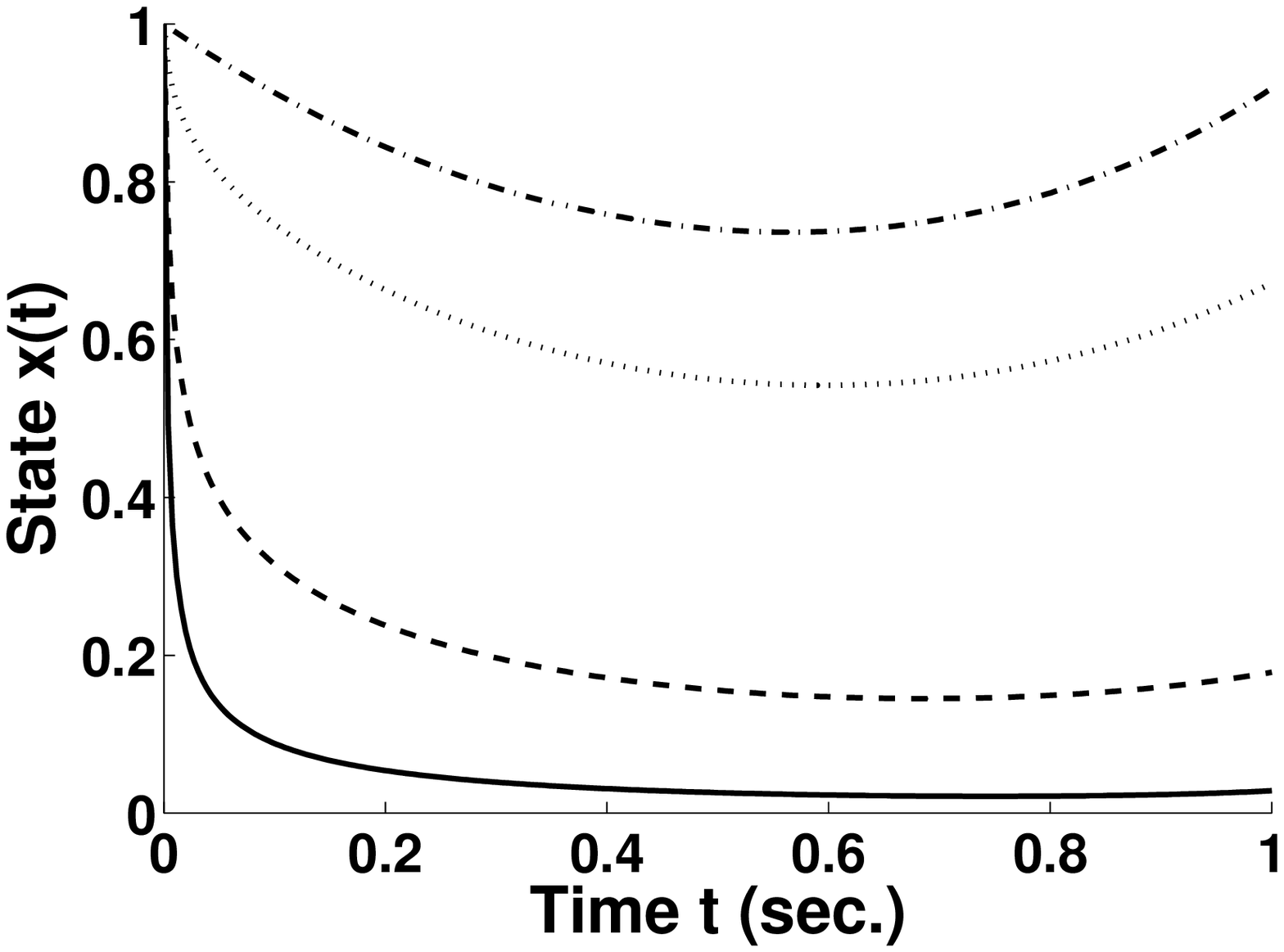}

Figure 11: State $x(t)$ as a function of $t$ for the TVP for
different $\alpha$\\ ( $ -$ :$\alpha=0.5$, -$ $-$ $-
:$\alpha=0.75$, $\cdot$$\cdot$$\cdot$ :$\alpha=0.95$,

-$\cdot$-$\cdot$- :$\alpha=1$)
\end{center}

\begin{center}
\epsfxsize=9cm \epsfbox{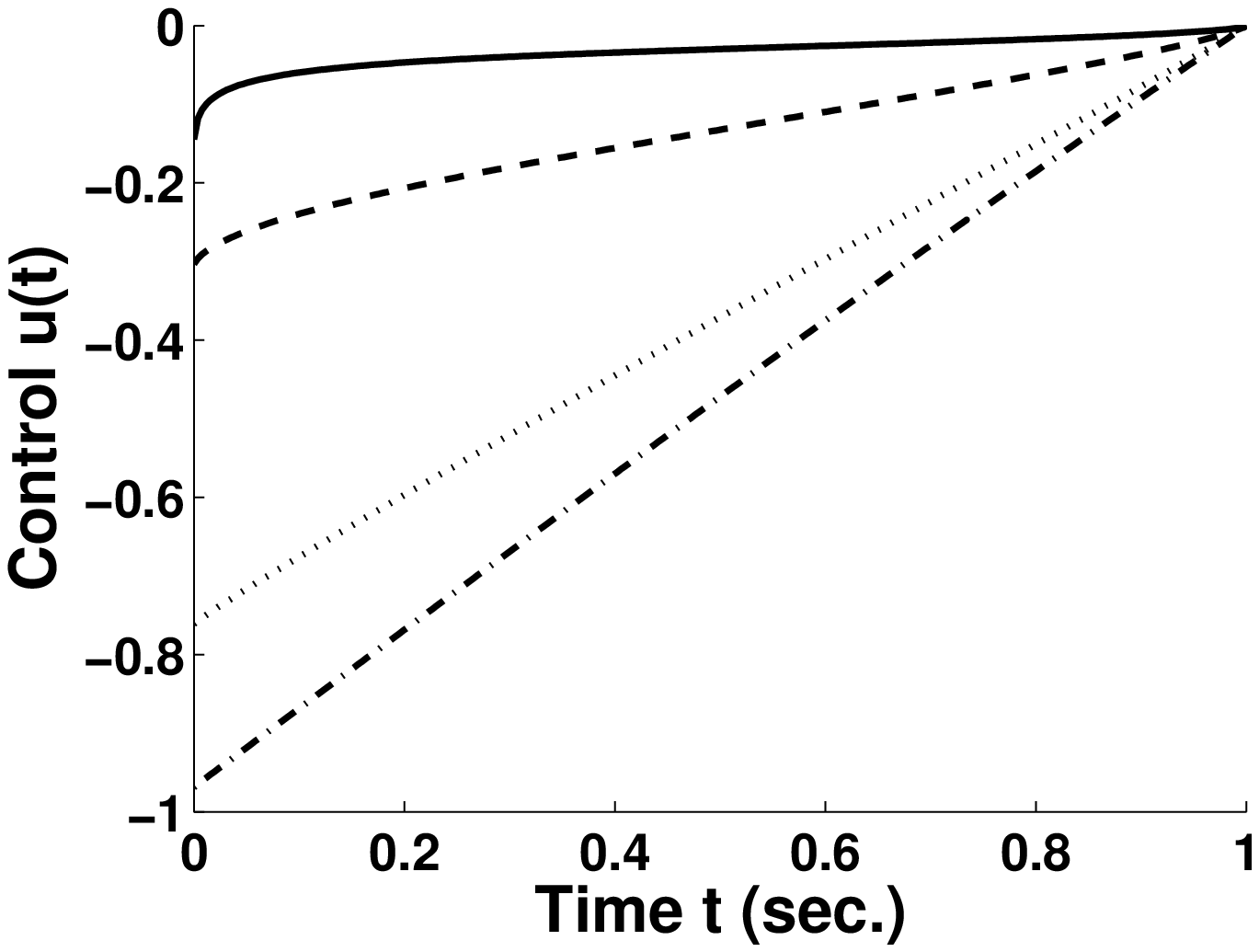}

Figure 12: Control $u(t)$ as a function of $t$ for the TVP for
different $\alpha$\\ ( $ -$ :$\alpha=0.5$, -$ $-$ $-
:$\alpha=0.75$, $\cdot$$\cdot$$\cdot$ :$\alpha=0.95$,
-$\cdot$-$\cdot$- :$\alpha=1$)
\end{center}

This problem for $\alpha=1$ has been solved in Agrawal (1989)
using a different scheme.The scheme is based on the approximation
with weighing coefficients and the lagrange multiplier technique
for a class of optimal control problems (Agrawal (1989)). Results
show that for $\alpha=1$ the numerical solutions obtained using
the scheme developed here and in Agrawal (1989) agree well. Thus,
as before, as $\alpha$ approaches to 1, the solution for the
integer order system is recovered.

It should be point out here that in integer order calculus central
difference schemes have been used in many cases to develop
numerically stable and efficient schemes.  It is hoped that this
research will initiate a similar effort in fractional calculus.

\section{Conclusions}

For a general class of fractional optimal control problems a
Hamiltonian was defined and a set of necessary conditions were
derived. A direct numerical scheme was presented for solution of
the problems. The scheme was used to solve two problems, time
invariant and time varying.  Results showed that as the number of
divisions of the time domain was increased, the solutions
converged. However, the convergence appears to be slow. As the
value of $\alpha$ approaches $1$, the solution for the integer
order system is recovered.  It is hoped that this research would
initiate further research in the field, and more efficient and
stable schemes would be found.

\section{Acknowledgments}

{ \small This work is partially supported by the Scientific and
Technical Research Council of Turkey.}\\

% ------------------------------------------------------------------------

\end{document}